\documentclass[default]{aastex701}

\usepackage{soul, xcolor}

\begin{document}

\title{Dynamical environment and stability around Centaur (2060) Chiron}

\author[orcid=0000-0001-5138-230X]{G.~Madeira}
\affiliation{Observat\'orio Nacional/MCTI, R. General Jos\'e Cristino 77, CEP 20921-400, RJ, Brazil}
\email[show]{madeira@on.br}

\author[0000-0003-0088-1808]{B.~E.~Morgado}
\affiliation{Universidade Federal do Rio de Janeiro - Observat\'orio do Valongo, Ladeira Pedro Ant\^onio 43, CEP 20080-090, RJ, Brazil}
\affiliation{Laborat\'orio Interinstitucional de e-Astronomia/LIneA, Av. Pastor Martin Luther King Jr, 126, Torre 3000 / sala 817. CEP 20765-000, RJ, Brazil}
\email{bmorgado@ov.ufrj.br}

\author[0000-0003-1000-8113]{C.~L.~Pereira}
\affiliation{Observat\'orio Nacional/MCTI, R. General Jos\'e Cristino 77, CEP 20921-400, RJ, Brazil}
\affiliation{Laborat\'orio Interinstitucional de e-Astronomia/LIneA, Av. Pastor Martin Luther King Jr, 126, Torre 3000 / sala 817. CEP 20765-000, RJ, Brazil}
\email{chrystianpereira@on.br}

\author[orcid=0009-0008-2716-2794,sname= Ramon]{G. Ramon}
\affiliation{UNESP - São Paulo State University, Grupo de Dinâmica Orbital e Planetologia, Av. Ariberto Pereira da Cunha, 333, Guaratinguetá, 12516-410, SP, Brazil}
\email{}

\author[orcid=0000-0002-4939-013X,sname=Sfair]{R. Sfair}
\email{rafael.sfair@unesp.br}
\affiliation{UNESP - São Paulo State University, Grupo de Dinâmica Orbital e Planetologia, Av. Ariberto Pereira da Cunha, 333, Guaratinguetá, 12516-410, SP, Brazil}
\affiliation{LIRA, Observatoire de Paris, Université PSL, Sorbonne Université, Université Paris Cité, CY Cergy Paris Université, CNRS,  92190 Meudon, France}

\author[0000-0003-2311-2438]{F. Braga-Ribas}
\email{fribas@utfpr.edu.br}
\affiliation{Federal University of Technology - Paraná (PPGFA/UTFPR-Curitiba), Av. Sete de Setembro, 3165, CEP 80230-901 - Curitiba - PR - Brazil}
\affiliation{Laborat\'orio Interinstitucional de e-Astronomia/LIneA, Av. Pastor Martin Luther King Jr, 126, Torre 3000 / sala 817. CEP 20765-000, RJ, Brazil}
\affiliation{LIRA, Observatoire de Paris, Université PSL, Sorbonne Université, Université Paris Cité, CY Cergy Paris Université, CNRS,  92190 Meudon, France}



\begin{abstract}
A recent stellar occultation revealed that the Centaur (2060) Chiron hosts a broad disk extending beyond $\sim$200 km from its centre, embedding three ring-like structures (Chi1R, Chi2R, and Chi3R), while a tenuous outer ring (Chi4R) lies beyond the Roche limit. Here, we present a first dynamical assessment of the system’s stability through numerical simulations of test particles, accounting for Chiron’s triaxial figure. For an equatorial ellipticity of $C_{22}\sim0.02$, as inferred from the most recent shape estimates, our simulations reveal a chaotic inner zone extending to $\sim$260 km, where particle lifetimes reach up to a year, while particles beyond $\sim$260 km can remain stable for at least a decade. These results suggest that the innermost portion of the disk is ephemeral and can only persist if continuously replenished. For lower ellipticity values ($C_{22}\lesssim0.012$), however, the entire disk is located within the stable region, regardless of Chiron’s mass. Under the physical parameters currently available in the literature, Chi2R is possibly linked to the 1:3 spin–orbit resonance, while Chi1R cannot be linked to the 1:2 resonance, as previously proposed, since this resonance is unstable. Instead, Chi1R and Chi3R may be associated with the 2:5 and 1:5 spin-orbit resonances, respectively. Both the 1:3 and 1:5 resonances are bifurcated, generating chaotic zones that may explain the gap in Chi2R and the longitudinal asymmetry observed in Chi3R.
\end{abstract}


\section{Introduction} \label{sec:intro}
Discovered in 1977 \citep{Kowal1979}, (2060) Chiron has long attracted attention due to its hybrid nature, displaying characteristics of both asteroids and comets \citep{Jewitt2009}. Its diameter has been constrained to about 180–220 km from thermal modeling, stellar occultations, and ALMA observations \citep{Campins1994,Fornasier2013,Lellouch2017,BragaRibas2023}. Since shortly after discovery, Chiron has also exhibited cometary activity, with variations in absolute magnitude and the appearance of a coma at heliocentric distances beyond 11~au \citep{Bus1989,MeechBelton1990}. This dual identity, together with its intermediate orbit between Jupiter and Neptune, makes Chiron a key laboratory for understanding the physical evolution of small bodies in the outer Solar System.

Early stellar occultations provided the first direct probes of material around Chiron. Events in 1993 and 1994 revealed transient, narrow, jet-like features in multiple light curves \citep{Bus1996,Elliot1995}. For nearly two decades the nature of these structures remained unclear. In 2011 occultation detected from Hawaii revealed two symmetric secondary drops \citep{Ruprecht2015}. Their morphology closely resembled the ring-like structures previously identified around Chariklo \citep{BragaRibas2014}, suggesting the presence of a narrow ring system around Chiron \citep{Ortiz2015}. This reinterpretation provided a unifying framework for understanding both the jet-like features observed in the 1990s and the long-term photometric variability of the object.

\begin{figure}[ht!]
\centering
\includegraphics[trim={0 2cm 0 1cm},clip]{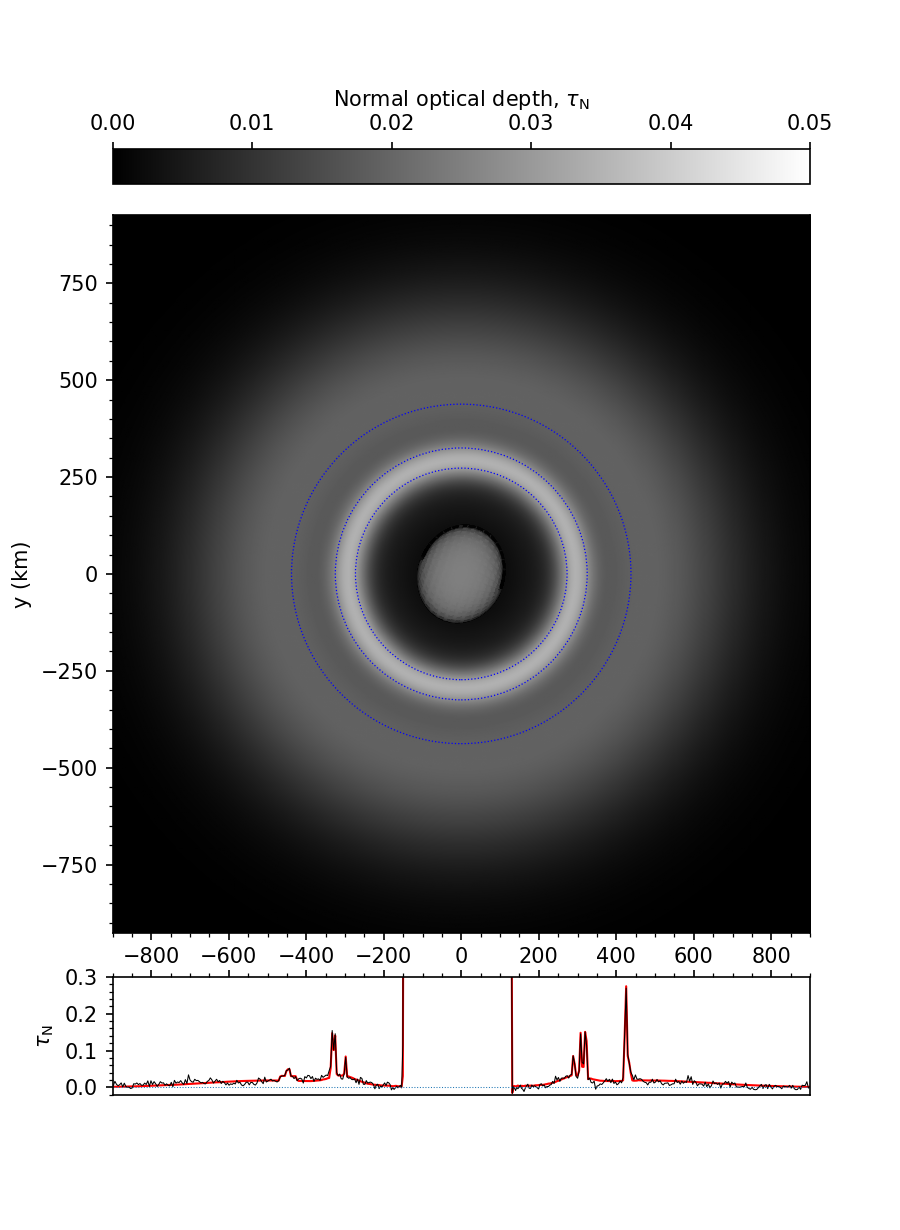}
\caption{Upper panel shows the ring plane pole-on view of the material around Chiron, with the grayscale proportional to the normal optical depth. The lower panel contains the normal optical depth curve obtained at Observatório do Pico dos Dias and the model fitted, including the ring and the disk material. The blue dotted lines represent the position of Chi1R, Chi2R, and Chi3R. Figure modified from \cite{Pereira2025}.}
\label{fig:OccModel}
\end{figure}

Subsequent occultations refined this picture. In 2018, ring-like structures were again detected around this Centaur \citep{Sickafoose2023}. The multi-chord event of 2019 constrained Chiron’s size and shape but revealed no additional structures, placing upper limits on the presence of broad material \citep{BragaRibas2023}. A breakthrough occurred in December 2022, when high-S/N observations from large telescopes revealed multiple secondary features at radii of $\sim$324, $\sim$423, and $\sim$580~km \citep{Ortiz2023}. 
Most recently, the September 2023 occultation revealed a broad disk-like structure with a radial extent of $\sim$550 km, embedding at least three dense rings (Chi1R, Chi2R, and Chi3R) and an additional tenuous outer feature (Chi4R) \citep{Pereira2025}. The rings exhibit optical depths and widths comparable to those of Chariklo \citep{BragaRibas2014, Berard2017, Morgado2021} and Quaoar \citep{Morgado2023, Pereira2023}, while the diffuse disk represents a new component not previously identified around small bodies. A concise view of the current material around Chiron can be seen in Figure \ref{fig:OccModel}. In a preliminary estimation, the innermost rings may coincide, within uncertainties, with the locations of the 1:2 and 1:3 spin–orbit resonances (SORs), while the more distant structures lie across the classical Roche boundary \citep{Pereira2025}.  

In other minor bodies with known ring systems -- Chariklo, Haumea, and Quaoar -- the shape of the central body plays a key role in shaping the dynamics of the surrounding material. Although Chiron’s figure remains poorly constrained, it has been reported to resemble a prolate ellipsoid with semi-axes of about $a_{\rm Ch}=126$ km, $b_{\rm Ch}=109$ km, and $c_{\rm Ch}=68$ km \citep{BragaRibas2023}. 
An ellipsoidal body generates gravitational perturbations that can expressed through spherical harmonics: 
the $C_{20}$ term represents oblateness (polar flattening), while the $C_{22}$  term represents prolate 
deformation or equatorial ellipticity.
Previous studies have shown that equatorial asymmetries strongly affect orbital eccentricities, efficiently clearing material from nearby regions \citep{Winter2019,Sicardy2020,Rollin2021,Madeira2022a,Ribeiro2023,GiuliattiWinter2023}.

Equatorial asymmetries also influence the locations of SORs and, consequently, ring dynamics. For Haumea, \citet{Winter2019,Ribeiro2023} show that when the body’s non-sphericity is taken into account, the reported ring location lies outside the 1:3 SOR, suggesting that the ring may not be confined by this resonance. In the case of Chariklo, \citet{Madeira2022a,GiuliattiWinter2023} likewise find that the main ring lies close to but outside the 1:3 SOR, both for a triaxial Chariklo and for a spherical Chariklo with a large mass anomaly. We point out, however, that these works neglect interactions between ring particles, which are expected to play an important role in the system’s stability \citep{Sickafoose2024}. Nonetheless, they consistently emphasize the central body’s figure as a key factor in shaping nearby particle dynamics and SOR locations, motivating us to investigate the role of Chiron’s shape in its circum-body environment.

In this work, we investigate the stability and dynamics around Chiron and their implications for the evolution of the disk and ring system detected around the object. In Section~\ref{sec:methods}, we describe our numerical simulations and in Section~\ref{sec:stability}, we present a general view of the dynamical environment around Chiron. Given the uncertainties in the object’s mass and shape, Section~\ref{sec:varying} explores how these parameters affect the system’s stability. Finally, in Section~\ref{sec:conclusion}, we present our conclusions.

\section{Methods} \label{sec:methods}

We perform our dynamical investigation with the \texttt{Rebound} package \citep{Rein2012}, using the IAS15 integrator \citep{Rein2015} to compute the trajectories of massless, non-interacting particles under Chiron’s gravitational potential. Chiron is taken as a triaxial ellipsoid, and in the quasi-inertial \texttt{Rebound} reference frame centered on the body ($x$, $y$, $z$), the particle equations of motion ($\ddot{x}$, $\ddot{y}$, $\ddot{z}$) are given by \citep{Celletti2017}:
\begin{equation}
\ddot{x}=-\frac{GM}{r^3}x+\frac{GMR^2}{r^5}\left[C_{20}\left(\frac{3}{2}x-\frac{15}{2}\frac{x z^2}{r^2}\right)+C_{22}\left(6x\cos{2\omega t}+6y\sin{2\omega t}+\frac{15x}{r^2}\xi\right)\right],
\end{equation}
\begin{equation}
\ddot{y}=-\frac{GM}{r^3}y+\frac{GMR^2}{r^5}\left[C_{20}\left(\frac{3}{2}y-\frac{15}{2}\frac{y z^2}{r^2}\right)+C_{22}\left(6x\sin{2\omega t}-6y\cos{2\omega t}+\frac{15y}{r^2}\xi\right)\right],
\end{equation}
\begin{equation}
\ddot{z}=-\frac{GM}{r^3}z+\frac{GMR^2}{r^5}\left[C_{20}\left(\frac{9}{2}z-\frac{15}{2}\frac{z^3}{r^2}\right)+C_{22}\frac{15z}{r^2}\xi\right],
\end{equation}
where $G$ is the gravitational constant, $M$ is the mass of Chiron, $r=\sqrt{x^2+y^2+z^2}$ is the distance to Chiron’s centre, $\omega$ is the spin frequency ($\omega=2\pi/T$, with $T$ being the spin period), $t$ is the simulation time, and $\xi$ is an auxiliary quantity defined as
\begin{equation}
\xi=\left(y^2-x^2\right)\cos{2\omega t}-2xy\sin{2\omega t}.
\end{equation}
The mean physical radius $R$ and the gravitational coefficients $C_{20}$ and $C_{22}$ are derived from the semi-axes $a_{\rm Ch}$, $b_{\rm Ch}$, and $c_{\rm Ch}$ of its triaxial shape \citep{Balmino1994}:
\begin{equation}
R=(a_{\rm Ch}b_{\rm Ch}c_{\rm Ch})^{1/3},
\end{equation}
\begin{equation}
C_{20}=\frac{2c_{\rm Ch}^2-a_{\rm Ch}^2-b_{\rm Ch}^2}{10R^2},
\end{equation}
\begin{equation}
C_{22}=\frac{a_{\rm Ch}^2-b_{\rm Ch}^2}{20R^2}.
\end{equation}
In Section~\ref{sec:stability} we examine the stability around Chiron, adopting the most recent constraints on its shape and mass, hereafter referred to as our nominal parameters, while in Section~\ref{sec:varying} we investigate the effects of varying Chiron’s mass ($M$) and equatorial ellipticity ($C_{22}$). As nominal parameters, we assume ${\rm M=4.8\pm2.3 \times 10^{18}~kg}$, ${\rm a_{Ch}=126\pm22~km}$, ${\rm b_{Ch}=109\pm19~km}$, and ${\rm c_{Ch}=68\pm13~km}$ \citep{BragaRibas2023}. These values yield $R=97.75$km, $C_{20}=-0.1937$, and $C_{22}=0.0209$. All simulations span $10^4T$, where Chiron’s rotation period is taken as $T=5.917813$~h \citep{Marcialis1993}, covering a period of about 6.75 years.

\subsection{Poincaré Surface of Section} \label{subsec:poincare}

The Poincaré Surface of Section, or Poincaré map, is a powerful technique for investigating stability. Originally developed to analyse resonance regions and to distinguish between regular and chaotic trajectories in the restricted, planar three-body problem \cite[e.g.][]{Poincare1895,Jefferys1971,Winter1994a,Winter1994b}, it has more recently been applied to the motion of test particles around rotating central bodies with non-spherical mass distributions. In particular, it has been used to study the stability around Haumea and Chariklo, both non-spherical symmetric bodies hosting ring systems \citep{Winter2019,Madeira2022a,Ribeiro2023,GiuliattiWinter2023}.

The trajectory of a particle in planar motion is unequivocally determined by its position components ($x_r$, $y_r$) and velocity components ($\dot{x}_r$, $\dot{y}_r$) in a frame rotating with the central body’s spin period. In short, a Poincaré map provides a two-dimensional representation of such a trajectory, originally defined in a four-dimensional phase space. To build the map, the particle’s Jacobi constant $C_J$ -- an integral of motion related to the initial velocity -- is fixed, and a reference plane is defined, usually $y_r=0$. The trajectory is then integrated, and each crossing of the reference plane in the direction $\dot{y}_r<0$ is plotted on the $x_r$–$\dot{x}_r$ phase plane. Particles in stable, regular motion trace closed patterns in the Poincaré map, forming stability islands, whereas chaotic trajectories appear as irregularly scattered points.

In this work, we employ Poincaré maps to identify the central locations and widths of the spin–orbit resonances in the system. A particle trapped in an $m$:$m+j$ SOR -- with $m$ and $j$ integers -- produces $j$ stability islands in the map, which makes it easy to visually distinguish a resonant particle from one that is merely stable but non-resonant. In total, we run 116 cases for $6080~{\rm m^2/s^2} \leq C_J \leq 8010~{\rm m^2/s^2}$, where \citep{Hu2004}
\begin{equation}
C_J = \omega^2(x_r^2+y_r^2) + 2U(x_r,y_r) - \dot{x}_r^2 - \dot{y}_r^2, \label{eq_cj}
\end{equation}
with \citep{Hu2004}
\begin{equation}
U(x_r,y_r) = \frac{GM}{r}\left(1 - \left(\frac{R_e}{r}\right)^2\left[\frac{C_{20}}{2} - \frac{3C_{22}}{r^2}(x_r^2 - y_r^2)\right]\right).
\end{equation}

For each simulation, 450 particles are integrated: 150 equally spaced in $200~{\rm km} \leq x_r \leq 700~{\rm km}$, each with $\dot{x}_r$ = 0, 3, and 6~${\rm m/s}$. We set initially $y_r=0$ and calculate the initial $\dot{y}_r$ from Eq.~\ref{eq_cj}. In post-processing, the particle trajectories are converted to the frame corotating with Chiron’s spin, and the libration angles of the main nearby resonances (1:1, 2:3, 1:2, 2:5, 1:3, 1:4, 1:5 SORs) are computed, identifying as resonant those particles with libration amplitudes below $180^\circ$ \citep[see][]{Sicardy2019,Madeira2022a,Ribeiro2023}. For simulations containing resonant particles, Poincaré maps are created by recording $(x_r,\dot{x}_r)$ at each crossing of $y_r=0$ with $\dot{y}<0$, linearly interpolating to obtain exact values. Finally, we visually identify the centres and edges of the resonant stability islands to determine the central locations and widths of the resonances, respectively. The $(x_r,\dot{x}_r)$ pairs identified in the Poincaré maps are then converted into semi-major axis and eccentricity \citep[for details, see][]{Madeira2022a}.

\section{Stability around Chiron} \label{sec:stability}

\begin{figure}[ht!]
\plotone{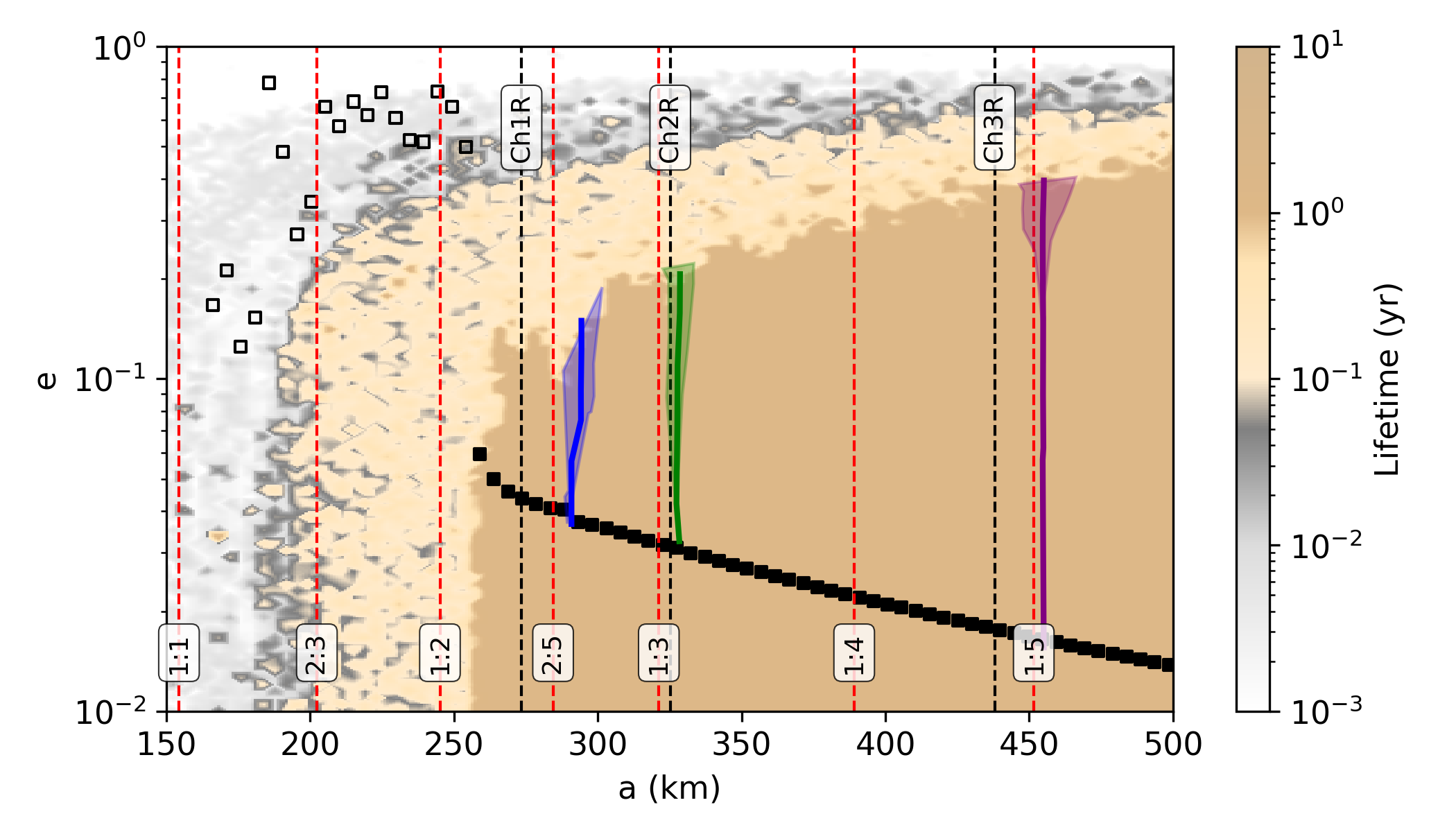}
\caption{Lifetime of particles in the vicinity of Chiron, assuming the object’s nominal parameters. Each colored pixel corresponds to a particle with a given initial semi-major axis (x-axis) and eccentricity (y-axis), with color indicating its lifetime. Filled black squares mark the maximum eccentricity recorded for particles that begin on Keplerian circular orbits and survive until the end of the simulation, whereas open black squares indicate the last values of the eccentricities registered by the N-body code for particles that are lost through ejection or collision with Chiron. Vertical red lines mark the theoretical locations of spin–orbit resonances, colored curves the resonance centers extracted from Poincaré maps, and shaded regions their widths. The observed ring locations are shown by vertical black lines.}
\label{fig:nominalparameters}
\end{figure}

We begin our study by exploring Chiron with its nominal parameters, that is, adopting the mass and shape model reported by \citet{BragaRibas2023}. As consistently shown in previous works \citep{Sicardy2020,Rollin2021,Madeira2022a,Ribeiro2023,beauge2025}, equatorial asymmetries in the central body induces variations in the eccentricities of nearby particles. Therefore, as a first test to evaluate the effect of Chiron’s mass distribution, we perform simulations with 100 particles uniformly distributed at semi-major axes ranging from 160 to 500 km compassing all three rings. The initial eccentricity, inclination, and angular orbital elements of all particles are set to zero.

Due to a combination of close encounters with Chiron’s equatorial bulge and the strong overlap of first-order SORs (m:m+1, for ${\rm m\geq3}$) near the synchronous radius, located at $\sim$154 km, particles with semi-major axes within $\sim$260 km undergo chaotic eccentricity diffusion \citep[for details, see][]{Rollin2021} and are eventually removed from the system, either through collisions with Chiron or by reaching hyperbolic orbits. In Figure~\ref{fig:nominalparameters}, the last values of the eccentricities registered by \texttt{REBOUND} for such lost particles are shown as open squares. In contrast, particles with semi-major axes greater than 260 km reside in a region without overlapping first-order SORs, where Chiron’s shape excites only a maximum eccentricity of $\sim10^{-2}-6\times10^{-2}$ -- as shown by the filled squares in Figure~\ref{fig:nominalparameters} -- indicating the existence of a stable region beyond this limit.

For a complete evaluation of the stability around Chiron, we create a grid of $300 \times 300$ particles, with semi-major axes uniformly distributed between 150 km and 500 km, and eccentricities logarithmically spaced between $10^{-2}$ and 1. The lifetimes of the particles are shown by the color-map in Figure~\ref{fig:nominalparameters}, where three distinct regions clearly emerge.

In the first one (gray region), radially extending up to $\sim$200 km for lower eccentricities, the particle pericenter lies within the strong overlap of first-order SORs, and collisions with Chiron typically occur in less than a month. In the second region, extending from $\sim$200 km to $\sim$260 km (light brown), where only the 2:3 and 1:2 first-order SORs are present, particles generally reach hyperbolic orbits on timescales of a few months to about a year. Beyond this region (dark brown), particles survive for at least several years.

If the nominal values are correct, an immediate implication is that the innermost portion of the observed disk, extending beyond $\sim$200 km, is likely ephemeral and therefore recent, since it should be efficiently cleared by Chiron’s shape up to the stable region. Nonetheless, the last outburst documented to Chiron happened between February and June of 2021 \citep{Ortiz2023, Dobson2024} and the occultation with the detection of Chiron's ring and disk of material happened in September 2023 \citep{Pereira2025}, which means, if the outburst originated the disk of material, the material should be able to survive for at least a few years. This indicates that an unaccounted-for efficient source of material is replenishing the disk in this inner region, or that the stable region around Chiron, in fact, extends inward to within $\sim$200 km. We explore this latter hypothesis in Section~\ref{sec:varying}.

Chiron’s rings are all observed to reside within the stable region. Nonetheless, contrary to the initial proposal of \citet{Pereira2025}, it seems unlikely that Chi1R is associated with the 1:2 SOR, since this resonance lies in the unstable region. In fact, we do not find any stable particles in this resonance in our simulations. Comparing the theoretical locations of SORs (red dashed lines), obtained following \citet{Sicardy2020}, with the observed positions of the rings (black dashed lines), we find that Chi1R lies relatively close to the 2:5 SOR, suggesting a possible association. Similarly, Chi2R is located near the 1:3 SOR, while Chi3R is closer to the 1:5 SOR. This correspondence is stronger for Chi2R, but weaker for both Chi1R and Chi3R.

To better assess these matches, we determine the locations and widths of the SORs using Poincaré maps, since analytical estimates do not account for the effect of $C_{22}$ \citep{Ribeiro2023}. The 1:3 and 1:5 SORs, located farther from the body, remain close to their theoretical predictions. In particular, the 1:3 SOR reaches widths of up to 6 km, comparable to the mean width of each component of Chi2R ($\sim$4-5 km), suggesting a possible correlation with this ring. The 1:5 SOR attains a maximum width of 11 km, which falls within the observed width of Chi3R, ranging from 2.9 to 47.1 km. In contrast, the 2:5 SOR shows a marked displacement between its theoretical and numerical locations, as expected from the stronger influence of $C_{22}$ in this region. As a result, the actual position of the 2:5 SOR is shifted by $\sim$20 km outward from Chi1R, leaving its possible association unresolved.

\begin{figure}[ht!]
\plotone{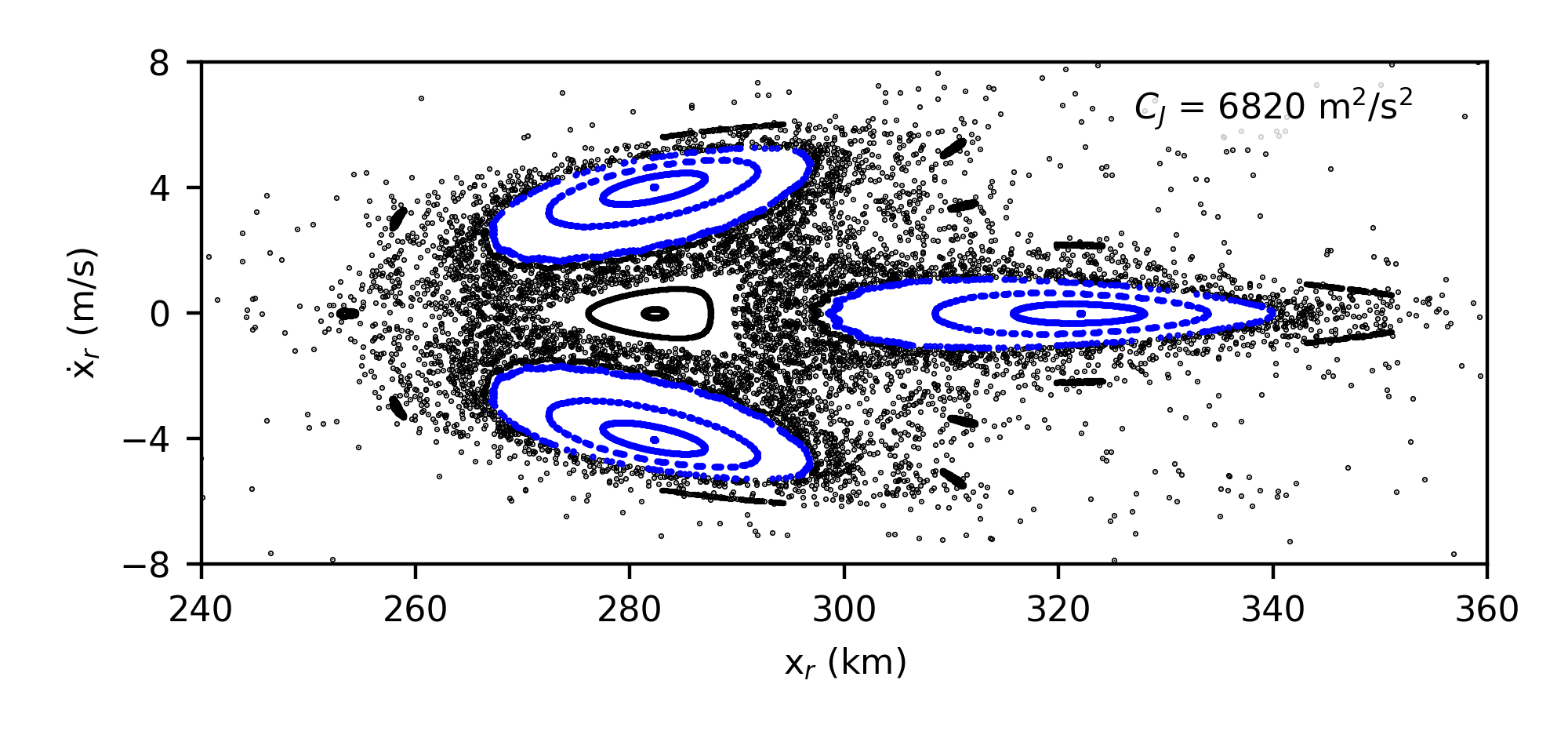}
\plotone{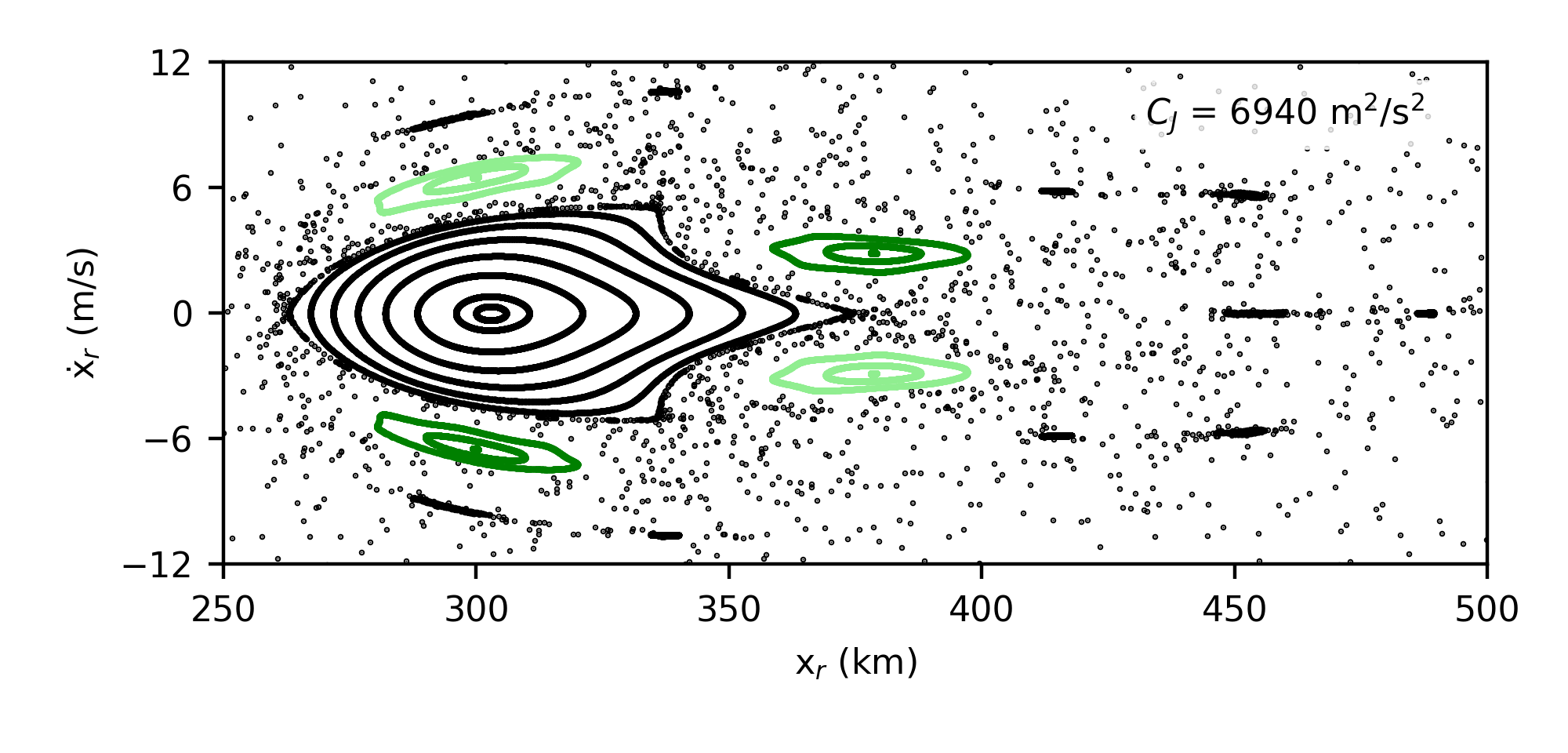}
\plotone{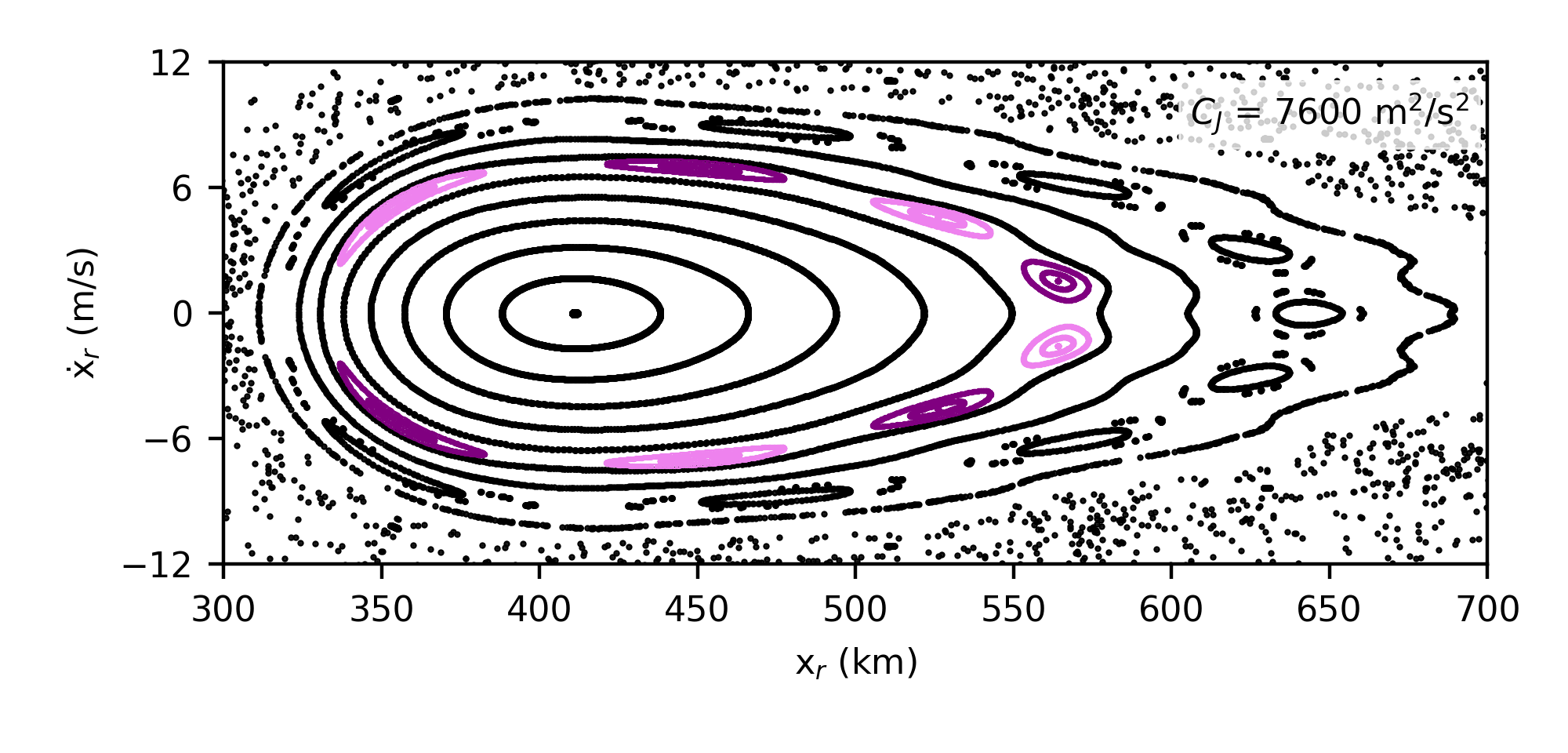}
\caption{Selected Poincaré maps showing the stability islands of spin–orbit resonances closest to each ring. The top panel highlights in blue the three islands of the 2:5 SOR, near Chi1R; the middle panel, in shades of green, the double islands of the 1:3 SOR, near Chi2R; and the bottom panel, in shades of purple, the double islands of the 1:5 SOR, near Chi3R. The Jacobi constant associated with each map is indicated in the upper-right corner of the panels.}
\label{fig:nominalparameters_ssp}
\end{figure}

Figure~\ref{fig:nominalparameters_ssp} shows selected Poincaré maps of the 2:5, 1:3, and 1:5 SORs (from top to bottom). The islands associated with each SOR are marked by colored points, while black dots correspond to chaotic or stable non-resonant particles. It is well known that SORs of the type $1$:$(1+p)$ around prolate bodies are bifurcated \citep{Winter2019,Sicardy2020,Ribeiro2023}, which is indeed verified for the 1:3 and 1:5 SORs around Chiron.

The bifurcation of the 1:3 SOR means that the resonance has four stable points instead of two, as would be expected for a second-order resonance -- formally speaking, this resonance corresponds to the 2:6 SOR -- with resonant particles librating around a pair of these points. Therefore, the 1:3 SOR is in fact composed of two independent families of resonant orbits, represented by different shades of green in Figure~\ref{fig:nominalparameters_ssp}. The same behavior is observed for the 1:5 SOR, where the two families are represented by different shades of purple in Figure~\ref{fig:nominalparameters_ssp}.

Besides doubling the number of stable points, the bifurcation also doubles the number of unstable points of the resonance, generating a chaotic zone in its vicinity. We therefore propose that the observed gap in Chi2R and the longitudinal asymmetry in Chi3R may be related to the bifurcation of the 1:3 and 1:5 SORs, respectively. Chi1R, in turn, does not show any special feature like its two outer companions, which would be expected if it is associated with the non-bifurcated 2:5 SOR, or if it is not related to any SOR at all.

\section{Evaluating Chiron's physical properties} \label{sec:varying}

Given the mismatch between the stable region obtained in Section~\ref{sec:stability} and the disk observed beyond $\sim$200 km around Chiron, we investigate how the location of the inner edge of the stable region depends on Chiron’s poorly constrained mass and shape. To this end, we simulate, for each pair of mass and $C_{22}$, 400 particles uniformly distributed in semi-major axis between 100 km and 500 km, with all other orbital elements set to zero. We then define the boundary of the stable region as the minimum semi-major axis beyond which particles survive. For larger Chiron masses ($\gtrsim7\times10^{18}$~kg), where the gravitational influence of the primary is stronger, some particles affected by the 1:1 and 1:2 SORs remain stable, whereas nearby particles do not. In these cases, we exclude such resonant particles and define the boundary as the minimum semi-major axis of the surviving particles located outside these resonances. This analysis is carried out on a uniform grid of $300 \times 300$ values of mass and $C_{22}$

Because Chiron’s dimensions are uncertain \citep{BragaRibas2023}, we explore a wide range of $C_{22}$ values, from 0 (corresponding to an oblate spheroid) up to 0.1. Regarding the mass, we have that the estimate of $4.8 \pm 2.3 \times 10^{18}$ kg proposed by \citet{BragaRibas2023} is model-dependent, relying on the assumption that Chiron is a fluid body in hydrostatic equilibrium with a Jacobi shape. To adopt a more conservative approach, we consider masses in the range $10^{18}$–$10^{19}$ kg, corresponding to bulk densities between $\sim$0.5 and $\sim$1.5 g/cm$^3$. This density range can be interpreted as variations in internal porosity, where lower bulk densities indicate higher macroporosity typical of rubble-pile structures, while higher densities suggest more consolidated, less porous material. Assuming typical grain densities of $\sim 2.5~\rm{g/cm^3}$ for carbonaceous material, these bulk densities correspond to distention parameter $\alpha \sim 0.2–0.6$, indicating porosities ranging from $\sim 40\%$ to $\sim80\%$.

\begin{figure}[ht!]
\plotone{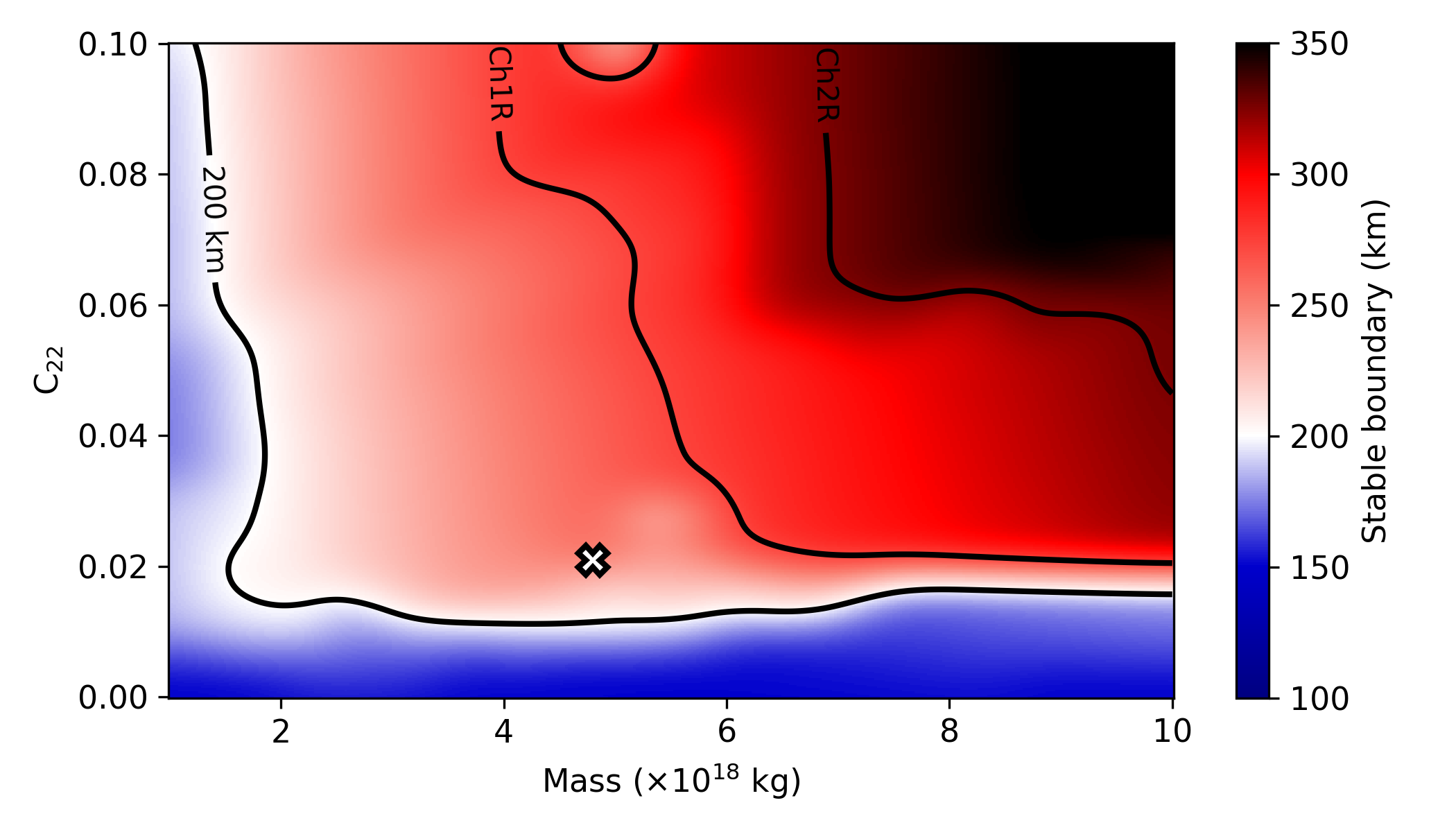}
\caption{Stable region boundary around Chiron as a function of its mass and $C_{22}$, shown by the colormap. Black curves mark the parameter combinations where the boundary coincides with the inner edge of the disk, Chi1R, and Chi2R. The white ‘X’ indicates Chiron’s nominal parameters.}
\label{fig:boundary_mass}
\end{figure}

The stable region boundary is shown in Figure~\ref{fig:boundary_mass}, with the black curves marking the combinations where this boundary coincides with the inner edge of the detected disk, Chi1R, and Chi2R. Larger values of $C_{22}$ correspond to greater equatorial ellipticity, leading to stronger excitation of particle eccentricities and thus expanding the extent of the unstable region. In the limiting case of an oblate spheroid, where Chiron has no ellipticity, no unstable region is detected around the body. Regarding the mass, a proportional relation between its value and the location of the stable boundary is also observed. Increasing the mass shifts the locations of the SORs responsible for chaotic eccentricity diffusion farther away from the central body, thereby broadening the chaotic region.

For the stable region to coincide with the observed disk, two solutions emerge. One possibility is that Chiron is less elongated, with $C_{22}\sim0.012$ already sufficient to reconcile the two, regardless of its mass. Alternatively, reconciliation for the nominal value of $C_{22}$ is achieved only if Chiron has a mass $\lesssim2\times10^{18}$~kg, implying a bulk density lower than 0.51~g/cm$^3$. Based on the parameters proposed by \cite{BragaRibas2023}, the difference in $C_{22}$ falls within the 1-sigma confidence level, making this scenario more likely than difference on Chiron's mass, which falls within the 3-sigma confidence level.

An alternative hypothesis is that the innermost portion of the disk is indeed ephemeral and will eventually be cleared, with the inner edge of Chi1R defined by the unstable region. This scenario, however, would require Chiron to have a $C_{22}$ value at least twice the nominal one, and possibly a higher mass, which appears unlikely since such values fall outside the 1-sigma confidence level for its figure. In addition, highly elongated bodies are expected to produce significant distortions in the positions of nearby SORs \citep{Ribeiro2023}, implying that Chi2R would likely not be located close to the 1:3 SOR in this scenario.

\begin{figure}[ht!]
\plotone{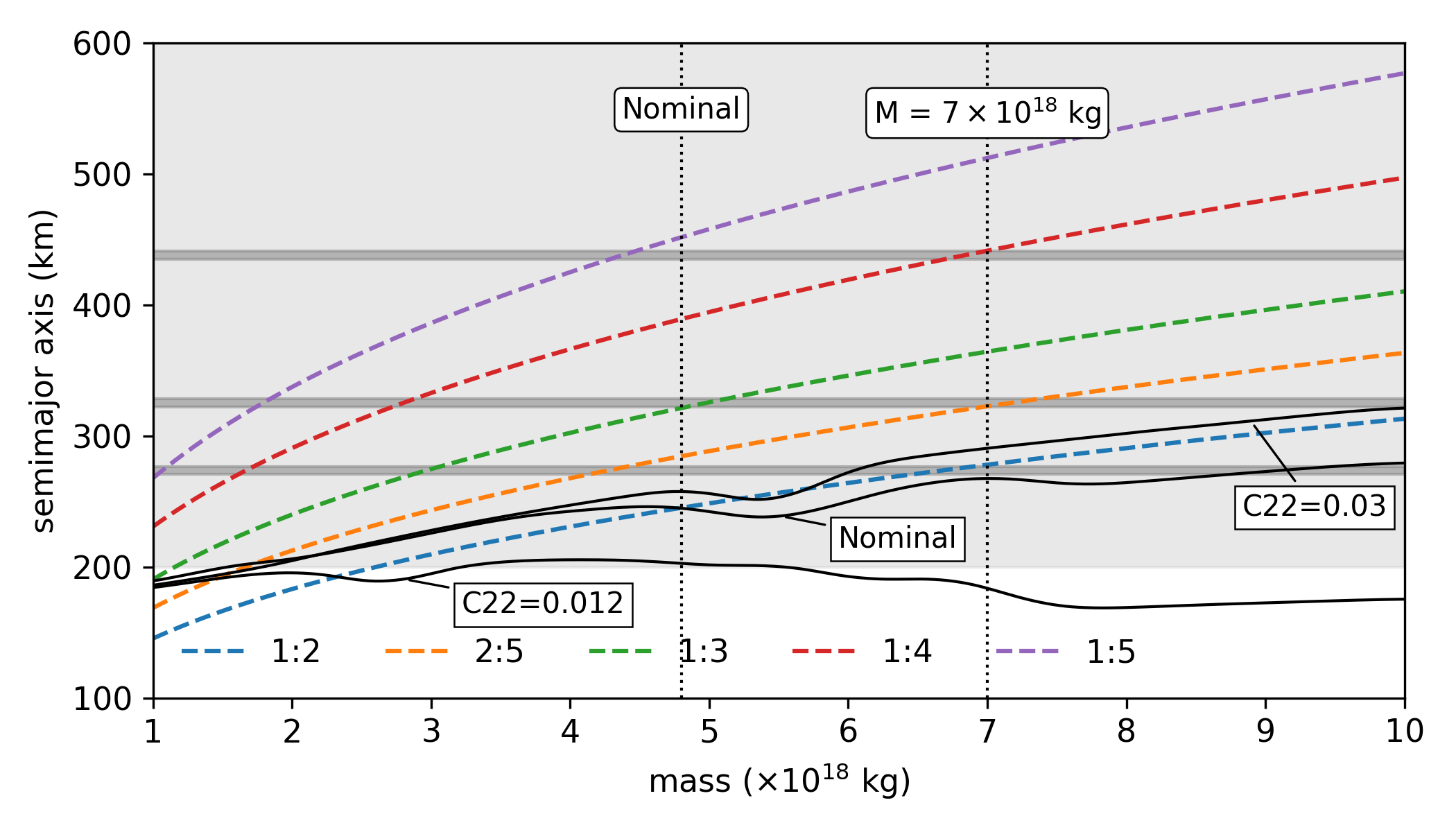}
\caption{Theoretical locations of spin–orbit resonances (dashed lines) and stable region boundaries (black curves) for different $C_{22}$ values, as a function of Chiron’s mass. The grey area marks the extent of the detected disk, and the narrow dark-grey bands indicate the rings.}
\label{fig:resso_loc}
\end{figure}

Besides affecting the extent of the stable region, Chiron’s physical properties also shift the locations of the SORs \citep{Sicardy2020}, which are envisioned to be associated with the rings. To evaluate this influence, we show in Figure~\ref{fig:resso_loc} the theoretical locations of the main SORs as a function of Chiron’s mass. While theoretical resonance locations depend only on the mass and the spin, the actual positions are known to depend also on $C_{22}$. We are therefore aware that this approach has limitations; nonetheless, for simplicity -- and because the previous section shows reasonable agreement between theoretical and numerical values -- we restrict the analysis to the theoretical predictions.

For any mass below $1.8\times10^{18}$~kg, the nominal $C_{22}$ provides a solution that simultaneously accounts for the disk and the stability of Chi1R and Chi2R -- which would be associated with the 1:4 and 1:5 SORs, respectively -- but leaves Chi3R without association with any resonance of order lower than four. Since SORs have been proposed as a possible explanation for rings’ positions and long-term stability, this point may argue against the hypothesis that Chiron is significantly less massive while maintaining the same figure. As a purely illustrative exercise, Figure~\ref{fig:resso_loc} marks the case of a Chiron with mass $7\times10^{18}$ kg, for which all three rings are simultaneously associated with SORs (1:2, 2:5, and 1:4, from inside to outside). However, this configuration fails to explain the long-term stability of the innermost part of the disk for $C_{22}\gtrsim0.012$, also requiring Chiron to be less prolate. 

At last, external effects such as collisions \citep{GilHutton2024, Cikota2018} or the removal of material due to outgassing \citep{PinillaAlonso2024} could have altered Chiron’s rotation period over the last decade. To test this possibility, we performed simulations to determine the stable region boundary on a uniform grid of $100 \times 100$ values of $C_{22}$ and spin period, the latter ranging from 3 to 8 hours. We acknowledge that such large changes in Chiron’s period are highly unrealistic, but we deliberately adopted this broad range to encompass even the most extreme scenarios. The results show that shorter periods bring the detected and stable regions closer together; however, for them to fully overlap, Chiron would need to rotate much faster than currently measured, with a spin period $\lesssim$3.2 h. This is close to the body’s critical rotation limit ($\sim$3 h for the nominal bulk density), making such a scenario extremely unlikely.

\section{Conclusion} \label{sec:conclusion}

The Centaur (2060) Chiron hosts a rich environment composed of a broad disk of particles spanning roughly 200–750 km, and four ring-like structures (Chi1R, Chi2R, Chi3R, and Chi4R) located at $\sim$273 km, $\sim$325 km, $\sim$438 km, and $\sim$1,380 km, respectively. Each of these features displays distinct properties, making Chiron a fascinating and challenging laboratory for dynamical studies. Here, we present a first glimpse of the system’s stability by investigating the dynamics of test particles around Chiron. 

Adopting the most recent estimates of the body’s mass and shape \citep{BragaRibas2023}, our simulations show that the region close to the object, extending radially up to $\sim$200 km, is strongly chaotic due to the overlap of first-order SORs, with particles surviving for no longer than a month. Beyond $\sim$200 km, where a disk of material has been detected around Chiron \citep{Pereira2025}, particles exhibit longer lifetimes. Between $\sim$200 km and $\sim$260 km, they typically survive from months to about a year before reaching hyperbolic orbits as a result of eccentricity diffusion induced by Chiron’s non-spherical shape. Further than $\sim$260 km, particles show stable motion, surviving for at least a decade. These results suggest that the innermost portion of the observed disk is ephemeral and unlikely to persist over long timescales without replenishment.

Even though there is no evidence of a possible outburst in 2023 \citep{Dobson2024}, a possible explanation for this discrepancy between the short dynamical lifetimes and the material detected in 2023, two years after the 2021 outburst is that additional, smaller-scale outbursts may have occurred without being detected, continuously replenishing the disk. Alternatively, other processes, such as micrometeoroid impacts, could provide fresh material to the inner regions. These mechanisms would account for the presence of particles on timescales longer than those predicted by our nominal stability analysis.

A more likely explanation, however, is that the entire disk structure resides within the stable region, which would occur if Chiron has a different mass or degree of prolateness. Our investigation shows that for lower equatorial ellipticity values the stable region shifts inward, and for $C_{22}\lesssim0.012$ the disk is fully contained within the stable region, regardless of Chiron’s mass. That is a plausible value which lies within the 1-sigma confidence level of Chiron's shape and size. The disk is also found to be fully contained within the stable region for Chiron models with densities of $\sim$0.5~g/cm$^3$, an extreme value that makes these solutions less plausible. Moreover, in such cases, at least one of the three innermost rings is not associated with a SOR.

For the nominal parameters, the location of Chi2R closely matches the actual position of the 1:3 SOR, while Chi1R and Chi3R can be associated with the 2:5 and 1:5 SORs, respectively. As previously found for other triaxial bodies \citep{Winter2019,GiuliattiWinter2023}, the 1:3 SOR in Chiron is also bifurcated, generating a chaotic zone in its vicinity. We envision that this chaotic region may be responsible for the gap detected in Chi2R. Similarly, the 1:5 SOR is also bifurcated, and its associated chaotic zone might be linked to the longitudinal asymmetry observed in Chi3R.

We highlight, however, that this is a first-order study that neglects interactions between ring particles. As in the rings observed around other minor bodies, Chiron’s rings are estimated to be relatively dense, implying that self-gravity and collisional dissipation are expected to strongly influence their stability \citep{Sickafoose2024,Sickafoose2025}. In particular, results from \citet{Sicardy_sub} indicate that only first- and second-order SORs can effectively confine ring structures, suggesting that Chiron’s rings may not, in fact, be directly associated with the SORs mentioned above, or at least that these resonances might not be responsible for maintaining the rings confined. Furthermore, our model neglects possible mass anomalies within Chiron, which could shift or even aid the confinement of the rings by SORs \citep{Sicardy_sub}. Future stellar occultations and improved modeling of Chiron’s shape are expected to provide tighter constraints, enabling a more complete characterization of its dynamical environment and the possible associations between the observed rings, the disk, and the nearby stable regions.

\begin{acknowledgments}
We thank Bruno Sicardy for valuable discussions and the anonymous reviewer for comments that helped improve this article. GM acknowledges the infrastructure provided by the Multiuser Data Processing Center of the National Observatory (CPDON). CLP thanks the FAPERJ/DSC-10 E-26/204.141/2022, FAPERJ/PDR-10 E-26/200.107/2025, and FAPERJ 200.108/ 2025. GR and RS are grateful to FAPESP, grant 2024/20150-1. RS and FBR thanks CNPq grants 307400/2025-5 and 316604/2023-2, respectively. The numerical simulations were performed on the CDJPAS platform at National Observatory, Brazil. During the preparation of this work, the authors used ChatGPT-5 in order to improve readability. After using this tool, the authors reviewed and edited the content as needed and take full responsibility for the content of the publication.
\end{acknowledgments}

\begin{contribution}
GM led the project conception, designed and performed the numerical simulations, analyzed the results, and wrote the manuscript. BEM and CLP contributed to the project conception, compared the simulation results with observational data, and co-wrote the manuscript. GR contributed by writing specific parts of the manuscript. RS and FBR provided suggestions and comments that improved the analysis and the manuscript.

\end{contribution}

\bibliography{sample701}{}
\bibliographystyle{aasjournalv7}



\end{document}